\documentclass[conference,a4paper]{IEEEtran}
\normalsize
\ifCLASSINFOpdf
\else
\fi

\usepackage{amsthm}
\usepackage{amsmath}
\usepackage[mathscr]{euscript}
\usepackage{amssymb}
\usepackage{graphicx}
\usepackage{esint}
\usepackage{amsfonts}
\usepackage{cite}
\usepackage{balance}
\usepackage{optidef}
\usepackage{caption}
\usepackage{subcaption}
\usepackage{epstopdf}
\usepackage{color}
\usepackage{bm}
\makeatletter
\usepackage{mathtools}
\usepackage{algpseudocode}
\usepackage{algorithm}
\usepackage{enumitem}
\usepackage{mathtools, nccmath}
\usepackage{float}
\theoremstyle{plain}
\newtheorem{thm}{\protect\theoremname}

\makeatother

\providecommand{\theoremname}{Theorem}

\theoremstyle{plain}

\makeatother

\providecommand{\lemmaname}{Lemma}

\theoremstyle{plain}

\makeatother

\providecommand{\remarkname}{Remark}

\DeclareSymbolFont{matha}{OML}{txmi}{m}{it}
\DeclareMathSymbol{\varv}{\mathord}{matha}{118}

\begin{document}




\title{Power-Optimal HARQ Protocol for Reliable Free Space Optical Communication \vspace{-0.4cm}}
\author{\IEEEauthorblockN{Georgios D. Chondrogiannis\IEEEauthorrefmark{1}, Nikos A. Mitsiou\IEEEauthorrefmark{1}, Nestor D. Chatzidiamantis\IEEEauthorrefmark{1},\\ Alexandros-Apostolos A. Boulogeorgos\IEEEauthorrefmark{2}, and George K. Karagiannidis\IEEEauthorrefmark{1}\IEEEauthorrefmark{3}
 } 
 \vspace{0.35mm}
\IEEEauthorblockA{\IEEEauthorrefmark{1}Department of Electrical
and Computer Engineering, Aristotle University of Thessaloniki,
GR-54124 Thessaloniki, Greece \\
 \vspace{0.35mm}
\IEEEauthorblockA{Emails: \{gchondro,nmitsiou,nestoras,geokarag\}@auth.gr} 
}
\IEEEauthorblockA{
\IEEEauthorrefmark{2}Department of Electrical and Computer Engineering, University of Western Macedonia, 50100 Kozani, Greece.\\
 \vspace{0.35mm}
\IEEEauthorblockA{Emails: al.boulogeorgos@ieee.org\\
\IEEEauthorrefmark{3}Cyber Security Systems and Applied AI Research Center, Lebanese American University (LAU), Lebanon \\
 \vspace{0.35mm}
\vspace{-0.3in}
}}

\vspace{-0.5cm}}
\maketitle

\begin{abstract}
This paper investigates the usage of hybrid automatic repeat request (HARQ) protocols for power-efficient and reliable communications over free space optical (FSO) links. By exploiting the large coherence time of the FSO channel, the proposed transmission schemes combat turbulence-induced fading by retransmitting the failed packets in the same coherence interval. To assess the performance of the presented HARQ technique, we extract a theoretical framework for the outage performance. In more detail, a closed-form expression for the outage probability (OP) is reported and an approximation for the high signal-to-noise ratio (SNR) region is extracted. Building upon the theoretical framework, we formulate a transmission power allocation problem throughout the retransmission rounds. This optimization problem is solved numerically through the use of an iterative algorithm. In addition, the average throughput of the HARQ schemes under consideration is examined. Simulation results validate the theoretical analysis under different turbulence conditions and demonstrate the performance improvement, in terms of both OP and throughput, of the proposed HARQ schemes compared to fixed transmit power HARQ benchmarks.


\end{abstract}

\section{Introduction}\label{S:Intro}

Free space optical (FSO) communication has emerged as a promising technology that can compensate for the growing scarcity of the radio-frequency spectrum. High-speed, wide bandwidth and low-cost wireless data transfer are some of the key benefits that FSO brings, despite the fact that the reliability of those systems has been a problem especially for long distances and foggy atmospheric conditions \cite{B:laser_beam}. As an enabler for ensuring an all-weather reliable FSO link over transmission distances of few kilometers, error-control retransmission protocols, that take into consideration the interconnection between the physical and the link layer, has been identified.

A recently proposed scheme for error-control retransmissions is the combination of automatic repeat request (ARQ) and forward error correction (FEC), the so-called hybrid ARQ (HARQ). Specifically, ARQ realizes the transmission of erroneously received packets based on the feedback from the receiver, while FEC accomplishes the correction of received data errors by adding a few redundant bits to the transmitted data. Thus, transmission reliability is improved, which simultaneously boosts the system's throughput \cite{HARQ_IR}.

Several HARQ retransmission schemes have been proposed in the context of FSO communications (e.g. \cite{survey} and the references therein). Among them, two types have attracted particular interest, namely, HARQ with code combining (HARQ-CC) and HARQ with incremental redundancy (HARQ-IR). In both, previously failed packets are stored and combined with subsequent retransmissions for decoding. Specifically, in HARQ-CC the same packet is retransmitted, while in HARQ-IR redundant information is incrementally transmitted in each  round. HARQ-CC and HARQ-IR schemes were investigated in \cite{Uysal}, where an outage analysis under turbulence induced fading was documented. A similar analysis was presented in \cite{Alouiniperf,improv}, where pointing and misalignment errors were also taken into account. Moreover, other performance metrics such as waiting and sojourn times, were investigated in \cite{touati}. Note that in all existing HARQ retransmission protocols, any constraints regarding the power consumption that may exist at the FSO transmitter are not considered. Furthermore, it is assumed that each transmission attempt uses equal transmit power. 

However, the assumption of unlimited power consumption at the transmitter is not always valid. For example, energy restriction issues may arise when FSO links are established between a fixed-point station and moving entities, such as unmanned aerial vehicles, High-altitude platform stations (HAPS) or satellites \cite{das}. In the view of the above, we present two energy efficient HARQ retransmission protocols that modify the parameters in each transmission round such that the outage probability (OP) is minimized, while 
 reducing the power-consumption at the transmitter's side as well. In particular, the contribution of the paper is summarized below:
\begin{itemize}[leftmargin=2.5mm]
    \item We investigate the performance of HARQ-CC and HARQ-IR protocols, under the assumption that the use of interleaving is impractical in FSO links  due to their large coherence times \cite{pointing}. Closed-form,  analytical expressions are derived for the OP of each protocol, while taking into account both atmospheric and misalignment effects. 
    \item We further elaborate on the case of high signal-to-noise ratio (SNR) regime and derive tractable and insightful expressions for the outage performance of both protocols. 
    \item Based on the high SNR OP analysis, we address the problem of minimizing the OP and the average throughput subject to average and peak optical power constraints. Due to its non-convexity, a tractable solution is proposed, via the concept of successive convex approximation (SCA).
    \item Simulation results validate the theoretical analysis for the OP of both HARQ-CC and HARQ-IR. Also, optimized HARQ-CC and optimized HARQ-IR are shown to outperform their respective fixed power HARQ benchmarks, both in terms of OP and throughput. 
\end{itemize} 
\section{System model}\label{sec:SSM}
We consider a point-to-point intensity modulation-direct detection (IM/DD) FSO system where HARQ transmission schemes are employed. The received signal in the $i$-th transmission round can be expressed as
\begin{align}
	\mathbf{y}_{i}= \mathcal{R} P_{i} h_i \mathbf{x}_{i}+\mathbf{n}_{i},\,\text{with} \,\,\, i=1,2, \ldots ,J
 \label{eq:1}
\end{align}\noindent
where $\mathbf{x}_i$ is the $i$-th modulated optical signal with unitary mean, i.e. $\mathbb{E}[\mathbf{x}_i]\!=\!1$, $h_i$ stands for the $i$-th channel coefficient that models the fading process and $\mathbf{n}_i$ represents the zero-mean Gaussian noise at the receiver site at the $i$-th transmission round with variance $\sigma_{n}^2$, i.e., $\mathbf{n}_i\sim N (0,\sigma_{n}^2)$. Furthermore, $\mathcal{R}$ is the receiver's responsivity and $P_i$ is the optical transmit power that satisfies $P_i\leq P_{max}$ imposed by safety and physical limitations.

\begin{figure}[t]
\centering
	\includegraphics[height = 2.0cm, width = 8.6cm]{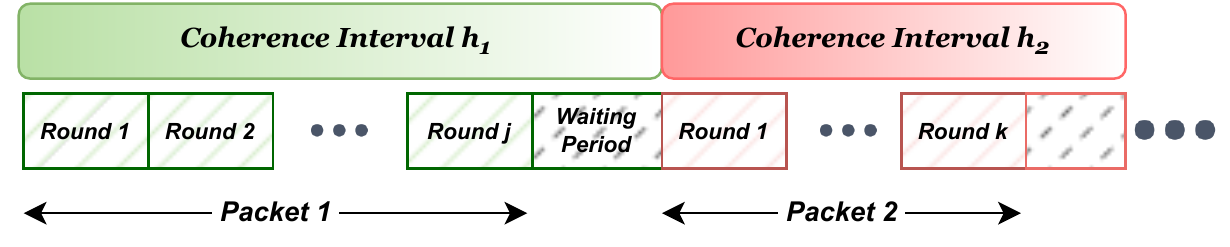}
	\caption{Relation between coherence intervals and packet transmission duration.}
	\label{Fig:coh}
    \vspace{-0.5cm}
\end{figure}

It should be noted that the optical channel results in a
very slowly-varying fading in FSO systems. For the signalling
rates of interest ranging from hundreds to thousands of Mbps
\cite{B:advanced}, turbulence-induced fading can be considered constant
over hundred of thousand or millions of consecutive symbols,
since the coherence time of the channel is about 1-100\,ms \cite{B:laser_beam}.
This fact leads to consider interleaving an unviable solution for averaging a large number of fading states, since the demands in storage memory would be unrealistic.
In the proposed transmission model, as depicted in Fig. \ref{Fig:coh}, each packet consists of several transmission rounds (at most $J$) and multiple transmitted symbols. Every packet experiences a different fading realization, where the independence between them is ensured through an idle waiting period.

In the analysis that follows, it is assumed that the
transmitter has statistical knowledge of the channel gains, i.e., statistical channel state information (CSI). This assumption is practical as it does not require extensive feedback from the receiver
and suffices to know only the statistical parameters of the
fading distribution (e.g. the probability density function of the SNR).

\subsection{Channel model}
The channel fading state, $h$, is considered to be the product of three factors, i.e., $h = h_l h_s h_g$. 
The first term, $h_l$, denotes the deterministic path loss exponent, which depends from link distance and weather conditions.
The second term, $h_g$, corresponds to the misalignment loss between the transmitter (TX) and the receiver (RX), i.e., pointing errors, of the optical point-to-point channel. Assuming a Gaussian beam profile at the receiver, the attenuation due to geometric spread with radial displacement from the origin of the detector follows a Rayleigh distribution. Finally, the third term stands for the attenuation due to the atmospheric turbulence conditions, i.e., scintillation \cite{pointing}. We consider the widely used Gamma-Gamma atmospheric attenuation model, in order to include a wide range of weak to strong turbulence conditions \cite{B:laser_beam}.

By combining the above statistical models, the probability density function (PDF) of $h= h_l h_s h_g$ is given as \cite{theory}
\begin{align}
	f_{h}(h) &=\frac{\alpha \beta \xi^{2}}{A_{0} h_l\Gamma(\alpha) \Gamma(\beta)}
	\nonumber\\ & \times\mathrm{G}_{1,3}^{3,0}\left[\alpha \beta \frac{h}{A_{0} h_l} \mathrel{\bigg|} \begin{array}{c}
		\xi^{2} \\
		\xi^{2}-1, \alpha-1, \beta-1
	\end{array}\right],
\end{align}
where $\alpha$ and $\beta$ are the statistical parameters that define the atmospheric turbulence conditions of the optical link \cite{B:laser_beam}. Furthermore, $\xi$ is the ratio between the equivalent beam radius at the receiver and the pointing error misplacement standard deviation at the receiver site ($\xi \rightarrow \infty$ corresponds to the case where there's no pointing errors). $A_0$ denotes a constant term of the pointing loss which represents the power collected at the detector's center. Finally, $G_{p, q}^{m, n}[\cdot]$ is the Meijer's G-function \cite[Eq. (9.301)]{B:grads}

\vspace{-0.2cm}
\subsection{Instantaneous SNR Statistics}
From \eqref{eq:1}, the instantaneous electrical SNR of the received signal in the $i^{\text{th}}$ transmission round is defined as
\begin{equation}
	\gamma_{i} = \frac{\mathcal{R}^2 P_i^2 h_i^2}{\sigma_n^2}.
        \label{eq:3}
\end{equation}
Based on \eqref{eq:3}, the cumulative distribution function
(CDF) of the instantaneous SNR are derived as \cite{unified}
\begin{align}
	F_{\gamma_i}(\gamma_i)=\frac{\xi^2}{\Gamma(\alpha) \Gamma(\beta)} G_{2,4}^{3,1}\left[\frac{\alpha \beta}{P_i}\sqrt{\frac{\gamma_i}{\bar{\gamma}}} \mathrel{\bigg|} \begin{array}{l}
		1, \xi^2+1 \\
		\xi^2, \alpha, \beta, 0
	\end{array}\right],
	\label{eq:9}
\end{align}
respectively, where 
\begin{align}
    \bar{\gamma}=\frac{A_0^2 h_l^2 \xi^2}{\sigma_n^2\left(\xi^2+1\right)} \stackrel{\xi^2 \gg 1}{=} \frac{A_0^2 h_l^2}{\sigma_{n}^2}
\end{align}
is the average received electrical SNR. Without loss of generality the receiver's responsivity is assumed to be unitary. Also the fading coefficient is given by $h_i = (A_0 h_l / P_i) \sqrt{\gamma_{i}/\bar{\gamma}}$.

\section{HARQ transmission protocols\label{sec:HARQ}}
\vspace{-0.2cm}
In this section, we investigate the performance of H-ARQ transmission protocols that can be applied in FSO systems and present optimal power allocation design strategies among transmission rounds. It is assumed that one-bit messages of positive or negative acknowledgement (ACK or NACK, respectively) are exchanged between TX and RX through a reliable and a zero-delay feedback channel.

\subsection{Code Combining H-ARQ}
In HARQ-CC, the TX sends the same codeword until an ACK is received or the 
maximum number of retransmissions $J$ is reached. At the RX's end, all the received copies of the encoded packet are combined using maximal ratio combining (MRC) and then the decoding method (e.g. Maximum likelihood) is performed. If the decoding attempt is successful an ACK message is sent back to the TX. 

By the end of the $j$-th transmission round, the accumulated mutual information is equal with  \cite{freespacecap,panosca} 
\begin{align}
I_{j}^{\mathrm{\textsc{cc}}}=\frac{1}{2j} \log _{2}\left(1+  c\sum_{i=1}^{j} \gamma_{i}\right),
\end{align}\noindent
where $c= \frac{1}{2\pi e}$ in IM/DD, if both peak-power and average-power constraints are imposed on the transmitted signal. By defining OP in $j$-th transmission round as the probability of the accumulated mutual information being smaller than the transmission rate $R$, it follows that  
\begin{align}
 &\mathbb{P}_{\text {out},j}^{\mathrm{\textsc{cc}}}\left(\mathbf{P}, R\right)
 =\Pr\left\{I_{j}^{\mathrm{\textsc{cc}}}\leq \frac{R}{j}\right\}
 \nonumber\\ &=\Pr\left\{h^2 \leq \frac{A_0^2 h_l^2 \left(2^{2R}-1\right) }{{c\bar{\gamma}\sum_{i=1}^{j}P_{i}^2 }}\right\},
 \label{eq:11}
\end{align}\noindent
where $\mathbf{P} = \left(P_1,\dots,P_j\right)$ is the vector of the transmitted power across $j$ HARQ rounds. After using \eqref{eq:9} and some basic algebraic manipulations, OP is calculated by
\begin{align}
	\mathbb{P}_{\text {out},j}^{\mathrm{\textsc{cc}}} & \left(\mathbf{P}, R\right)=\frac{\xi^{2}}{\Gamma(\alpha) \Gamma(\beta)}
	\nonumber\\ &\times G_{2,4}^{3,1}\left[\alpha \beta \sqrt{\frac{2^{2R}-1}{c \bar{\gamma} \sum_{i=1}^{j}P_{i}^2}} \mathrel{\Bigg|} \begin{array}{l}
		1, \xi^{2}+1 \\
		\xi^{2}, \alpha, \beta, 0
	\end{array}\right].
\end{align}
\subsubsection{High-SNR analysis}
In order to gain insights about the behavior of the proposed system the asymptotic analysis for the two protocols is performed.

\begin{thm}
In high-SNR region, the OP is approximated as 
\begin{equation}
	\mathbb{P}_{\text {out},j}^{\mathrm{\textsc{cc}}} \!\stackrel{\gamma_{i} \gg 1}{\cong}\! \begin{cases}
		\frac{\Gamma\left(\alpha-\xi^2\right)\Gamma\left(\beta - \xi^2 \right)}{\Gamma\left(\alpha\right)\Gamma\left(\beta\right)} \left(\frac{V_R}{\sum_{i=1}^{j} P_i^2}\right)^{\xi^2/2}, &\!\!\!\!\!\,\, \xi^2 < q\left(\alpha,\beta\right)
		\\
		\\
		C\left(\alpha,\beta\right) \left(\frac{V_R}{\sum_{i=1}^{j} P_i^2}\right)^{q\left(\alpha,\beta\right)/2},&\!\!\!\!\! \,\, \xi^2>q\left(\alpha,\beta\right)
	\end{cases}
\end{equation}
where $q\left(\alpha,\beta\right) = \min\{\alpha,\beta\}$, $V_R = \alpha^2 \beta^2 \left(2^{2R}-1\right)/c\bar{\gamma}$ and $C\left(\alpha,\beta\right) \!=\! \frac{\Gamma\left(\lvert\beta-\alpha\rvert\right)}{\left(1-q\left(\alpha,\beta\right)/\xi^2\right)\Gamma\left(q\left(\alpha,\beta\right)+1\right)\Gamma\left(\alpha\beta/q\left(\alpha,\beta\right)\right)}$.
\end{thm}
\begin{IEEEproof}
	The proof is provided in Appendix A.
\end{IEEEproof}

By defining  
\begin{align}
	\psi_R\left(\alpha,\beta\right) = \begin{cases}
		\frac{\Gamma\left(\alpha-\xi^2\right)\Gamma\left(\beta - \xi^2 \right)}{\Gamma\left(\alpha\right)\Gamma\left(\beta\right)} V_R^{\xi^2/2} & \text{for} \,\,
		\xi^2 < q\left(\alpha,\beta\right) \\ \\
		C\left(\alpha,\beta\right) V_R^{q\left(\alpha,\beta\right)/2} & \text{for} \,\, \xi^2 > q\left(\alpha,\beta\right)
	\end{cases}
\end{align}

\vspace{-0.2cm}
becomes evident that the asymptotic OP can be written as
\begin{align}
\mathbb{P}_{\text {out},j}^{\mathrm{\textsc{cc}}} \approx \frac{\psi_R\left(\alpha,\beta\right)}{\left(\sum_{i=1}^{j}P_{i}^2\right)^{\frac{\min\left(\xi^2,\alpha,\beta\right)}{2}}}.
\end{align}

\vspace{-0.1cm}
The average power across all rounds can be written as \cite{powerass}
\begin{align}
	\bar{P} = P_1 + \sum_{j=2}^{J} P_{j} \, \mathbb{P}_{\text {out,}j-1}^{\mathrm{\textsc{cc}}}.
\end{align}

\vspace{-0.1cm}
\subsubsection{HARQ-CC Optimization}
The problem addressed in this section is the optimal power allocation across the HARQ-CC rounds in order to minimize the OP subject to an average power constraint. To make the following optimization problem tractable we adopt the high-SNR analysis for the optical link. Thus, the optimization problem can be formulated as follows:

\begin{align}
	\nonumber\underset{P_1,P_2,...,P_J}{\mathbf{min}} \quad & \mathbb{P}_{\mathrm{out},J}^{\mathrm{{cc}}} \\
	\mathbf{s.t.} \quad & P_1 + \sum_{j=2}^{J}\frac{P_{j}\psi_R\left(\alpha,\beta\right)}{\left(\sum_{i=1}^{j-1}P_{i}^2\right)^{\frac{\min\left(\xi^2,\alpha,\beta\right)}{2}}} \leq P_0, \\ 
	\nonumber & 0\leq P_j\leq P_{\mathrm{max}},\quad \forall j \in \{1,...,J\}, \\ \nonumber
\end{align}

\vspace{-0.7cm}
where $P_0$ expresses the average power constraint and $P_\mathrm{max}$ denotes a peak power limitation for every round $j \in \{1,...J\}$. Note that the formulated problem is non-convex, due to both the imposed non-convex average power constraint and the objective function. As such, a viable solution cannot be provided. To that end, the following auxiliary variables $t_j$, $\forall j \in \{2,...,J\}$, are introduced for which it holds
\begin{equation}
    \frac{P_{j}\psi_R\left(\alpha,\beta\right)}{\left(\sum_{i=1}^{j-1}P_{i}^2\right)^{\frac{\min\left(\xi^2,\alpha,\beta\right)}{2}}} \leq t_j.
\end{equation}
Also, the auxiliary variables $\tilde{P_j}$ will be used, so that $\tilde{P_j}=P_j^2, \forall j \in \{1,...,J\}$. By using the properties of the natural logarithm, the optimization problem can be rewritten as 
\begin{align} \label{eq:semiconvex}
	\nonumber  \underset{P_1,P_2,...,P_J}{\mathbf{min}}& \quad  -\frac{\min\left(\xi^2,\alpha,\beta\right)}{2}\log\left(\sum_{j=1}^{J}\tilde{P}_{j}\right) \\
	\mathbf{s.t.} \quad & \sqrt{\tilde{P}_1} + \sum_{j=2}^{J} t_j \leq P_0 \\ 
	\nonumber & \log(\psi_R(a,b))+\frac{1}{2}\log(\tilde{P}_j)-\log(t_j)\\
	\nonumber& -\frac{\min\left(\xi^2,\alpha,\beta\right)}{2}\log\left(\sum_{i=1}^{j-1}\tilde{P}_{i}\right)\leq 0, \quad \\
	&\nonumber\forall j \in \left\{2,...,J\right\}\\
	\nonumber & 0\leq \tilde{P}_j\leq P_{\mathrm{max}}^2,\forall j \in \left\{1,...,J\right\}. \\ \nonumber
\end{align}

\vspace{-0.6cm}
The problem is still non-convex, due to the concave terms $\sqrt{\tilde{P}_1}$ and $\log(\tilde{P}_j)$ in the constraints. To overcome this, the concept of SCA is utilized. The first order Taylor expansion of the concave terms, around the randomly chosen initial point $\tilde{P}_{j,0}$, is given by
\begin{align} \label{eq:taylor}
&\sqrt{\tilde{P}_1}\approx \sqrt{\tilde{P}_{1,0}}+\frac{1}{2}\tilde{P}_{1,0}^{-\frac{1}{2}}(\tilde{P}_1-\tilde{P}_{1,0}), \\
&\log(\tilde{P}_j) \approx \log(\tilde{P}_{j,0})+\frac{\tilde{P}_{j}-\tilde{P}_{j,0}}{\tilde{P}_{j,0}}.
\end{align}
By substituting in \eqref{eq:semiconvex}, the problem which arises is convex. Thus, it can be solved by standard convex optimization methods, such as the interior-point method.
Moreover, due to the SCA, the solution of \eqref{eq:semiconvex} requires to solve its convex version multiple times, which is shown in Algorithm I.  
\begin{algorithm}
\begin{algorithmic}[1]\label{alg1}
\caption{HARQ-CC optimization} 
\State {\small{initialize FSO parameters, $\tilde{\mathbf{P}}_{0}$, $\delta_\mathrm{max}$ and $\epsilon$}}
\While {$\delta<\delta_\mathrm{max}$ AND $\rvert\rvert \mathbf{\tilde{P}}_\delta^*-\mathbf{\tilde{P}}_{\delta-1}^* \lvert\lvert_2^2>\epsilon$}
    \State{$\delta \gets \delta+1$}
    \State{Solve the convex version of \eqref{eq:semiconvex}, obtain $\mathbf{\tilde{P}}_\delta^*$}
    \State{$\mathbf{\tilde{P}}_0 \gets \mathbf{\tilde{P}}_\delta^*$}
\EndWhile
\State{$\mathbf{\tilde{P}}^*\gets \mathbf{\tilde{P}}_\delta^*$}
\end{algorithmic}
\end{algorithm}

\subsection{Incremental Redundancy H-ARQ}
In incremental redundancy protocol the source of the TX encodes the information message with a punctured version of a (low-rate) mother FEC code. At the initial transmission, only a few parity bits of the original code are transmitted along with the information message. If a decoding failure occurs, in the following rounds, the TX keeps sending additional redundant bits according to the puncturing pattern of the mother code.
At the other side, the RX combines the parity bits of the previous rounds with the most recently received ones and performs joint decoding. This particular scheme induces some extra complexity to the operation of both the TX and the RX, which comes as a consequence of the composite decoding process. 

\subsubsection{Performance Analysis}
The transmission rate for the first round, $R$, becomes $R/j$ after the passage of j rounds and the accumulated mutual information can be obtained as  
 \begin{align} 
	I_{j}^{\mathrm{\textsc{ir}}}
	&=\frac{1}{2j}\sum_{i=1}^{j}\log _{2}\left(1+ c\gamma_{i} \right),
\end{align}
and the OP is defined as 
\begin{align}
	\mathbb{P}_{\text {out},j}^{\mathrm{\textsc{ir}}}\left(\mathbf{P}, R\right)  =\Pr\left\{I_{j}^{\mathrm{\textsc{ir}}}\leq \frac{R}{j}\right\}.
\end{align}
We obtain an expression for the OP with the following approximation \cite{approx}.
\begin{align}
	\mathbb{P}_{\text {out},j}^{\mathrm{\textsc{ir}}}&\simeq\Pr\left\{\log_{2}\left(1+\left(c\bar{\gamma}\frac{h^2}{A_0^2 h_l^2}\right)^{j}\prod_{i=1}^{j} P_i^2\right)\leq 2R\right\}
	\nonumber\\&=\Pr\left\{h^2\leq \frac{A_0^2 h_l^2}{c\bar{\gamma}}\left( \frac{2^{2R}-1}{{\prod_{i=1}^{j}P_{i}^2 }}\right)^{1/j}\right\}
\end{align}

Similar with \eqref{eq:11}, the OP is given by 

\begin{align}
\nonumber \\ \noalign{\vskip-7pt}
\mathbb{P}_{\text {out},j}^{\mathrm{\textsc{ir}}}&\left(\mathbf{P}, R\right)=\frac{\xi^{2}}{\Gamma(\alpha) \Gamma(\beta)}
	\nonumber\\ &\times G_{2,4}^{3,1}\left[\frac{\alpha \beta}{\sqrt{c\bar{\gamma}}} \left(\frac{2^{2R}-1}{ \prod_{i=1}^{j}P_{i}^2}\right)^{1/2j} \mathrel{\Bigg|} \begin{array}{l}
		1, \xi^{2}+1 \\
		\xi^{2}, \alpha, \beta, 0
	\end{array}\right]
\end{align}

\subsubsection{High-SNR analysis}
Following the same methodology as in Section III, the OP can be asymptotically derived as:
\vspace{0.12cm}

\begin{align}
	\mathbb{P}_{\text {out},j}^{\mathrm{\textsc{cc}}} \approx \frac{\theta_{R,j}\left(\alpha,\beta\right)}{\left(\prod_{i=1}^{j}P_{i}\right)^{\frac{\min\left(\xi^2,\alpha,\beta\right)}{j}}}
\end{align}

where the nominator is given by 
\vspace{0.12cm}

\begin{align}
	\theta_{R,j}\left(\alpha,\beta\right) = \begin{cases}
		\frac{\Gamma\left(\alpha-\xi^2\right)\Gamma\left(\beta - \xi^2 \right)}{\Gamma\left(\alpha\right)\Gamma\left(\beta\right)} U_{R,j}^{\xi^2/2j} & \text{for} \,\,
		\xi^2 < q\left(\alpha,\beta\right) \\ \\
		C\left(\alpha,\beta\right) U_{R,j}^{q\left(\alpha,\beta\right)/2j} & \text{for} \,\, \xi^2 > q\left(\alpha,\beta\right)
	\end{cases}
\end{align}

\vspace{-0.2cm}
and $U_{R,j} = \left(\frac{\alpha^2\beta^2}{c\bar{\gamma}}\right)^{j}\left(2^{2R}-1\right)$
\\

\subsubsection{HARQ-IR Optimization}
In this section, the optimal power allocation across the HARQ-IR rounds is addressed. By applying a similar approach as in Section III, the optimization problem that arises is formulated as:
 \begin{align} \label{eq:IRopt}
 	\nonumber\underset{P_1,P_2,...,P_J}{\mathbf{min}} \quad & \mathbb{P}_{\text {out},J}^{\mathrm{\textsc{ir}}} \\
 	\textrm{s.t.} \quad & P_1 + \sum_{j=2}^{J}\frac{P_{j}\theta_{R,j}\left(\alpha,\beta\right)}{\left(\prod_{i=1}^{j-1}P_{i}\right)^{\frac{\min\left(\xi^2,\alpha,\beta\right)}{j}}} \leq P_0  \\ 
 	\nonumber&0\leq P_j\leq P_{\text{max}},\quad \forall j \in \{1,...J\} \\
 	\nonumber
 \end{align}

 \vspace{-0.4cm}
 The problem of \eqref{eq:IRopt} is nonconvex. However, it can be proven that \eqref{eq:IRopt} can be formulated as convex. Due to space limitation the analytic transformation is not given, nonetheless, by following a similar approach as in the HARQ-CC scheme \eqref{eq:IRopt} is transformed into a convex form. 

 \vspace{-0.1cm}
\subsection{Average Throughput}   
In order to gain a general perspective on the impact of power allocation to the FSO system, we investigate the throughput of the system throughout the implementation of the above techniques. The average throughput of HARQ schemes is defined as the average number of successfully delivered bits per channel use (bps/channel use)\cite{HARQ_IR}.
We are interested in finding the optimal power allocation $\mathbf{P^{\ast}}$ and the optimal value of the transmission rate $R$, for which the throughput is maximized. This can be described by the following optimization problem:
 \begin{align} \label{opt:eff}
\nonumber\underset{\mathbf{P^{\ast}},R}{\mathbf{max}} \quad \omega_J  = & \frac{R\left(1-\mathbb{P}_{\text{out}, J}\right)}{1+ \sum_{j=1}^{J-1} \mathbb{P}_{\text {out}, j}} \\
 	\textrm{s.t.} \quad & P_1 + \sum_{j=2}^{J} P_{j} \, \mathbb{P}_{\text {out,}j-1} \leq P_0  \\ 
 	\nonumber&0\leq P_j\leq P_{\text{max}},\quad \forall j \in \{1,...J\}, \\
 	\nonumber
 \end{align}

 \vspace{-0.4cm}
where $\mathbb{P}_{\text {out}, j}$ refers to either HARQ-CC or HARQ-IR protocol and the constraints are imposed from the assumed average and maximum optical power constraints of the FSO system. The problem is non-convex, however, it can be efficiently solved by utilizing the analysis of Section III, since, for a fixed value of $R$, maximizing the objective value of \eqref{opt:eff} is equivalent to maximizing $\left(1-\mathbb{P}_{\text{out}, J}\right)$. The intuitive reason for that is that $\left(1-\mathbb{P}_{\text{out}, J}\right)$ describes the probability of having a successful transmission at any round $j \in \{1,...,J\}$. Thus, maximizing that probability, also minimizes the OP of $\mathbb{P}_{\text{out}, j}, \forall j \in \{1,...,J\}$. Hence, by performing a linear search on $R$, problem \eqref{opt:eff} can be solved by finding the optimal solution of (14) for the HARQ-CC case, or (26) for the HARQ-IR case.


 \begin{table}[h]
 \resizebox{\columnwidth}{!}{%
 \begin{tabular}{ccc}
 	\hline Parameters & Symbol & Value \\
 	\hline  Average power constraint & $P_0$ &                  $200\mathrm{~mW}$ \\
        Maximum available power & $P_\mathrm{max}$ & $350\mathrm{~mW}$ \\
        Maximum retransmissions & $J $ & 4 \\
        Link range & $l$ & $1\mathrm{~km}$\\
        Receiver radius & $r$ & $10\mathrm{~cm}$\\
        Attenuation coefficient & $d_a$ & $\simeq 0.1$\\
        Ratio of EBR and the jitter & $\xi$ & $4$\\
 	Noise standard deviation & $\sigma_n$ & $10^{-7}\mathrm{~A}$ \\
        Jitter standard deviation & $\sigma_s$ & $30\mathrm{~cm}$\\
        \hline
        Algorithm 1 convergence threshold & $\epsilon$ & $10^{-5}$ \\
        Algorithm 1 iterations allowed   & $\delta_\mathrm{max}$ & 50 \\
 	\hline
 \end{tabular}
}
\end{table}

\vspace{-0.3cm}
\section{Results and Discussion\label{sec:results}}
In this section, simulations are presented to validate the presented performance analysis and also to demonstrate the effectiveness of the presented power allocation optimization problems. Unless otherwise stated, the simulation parameters are given in Table I. 
For the case of moderate turbulence we assumed $\alpha= 2.296$, $\beta = 1.822$, while for the case of strong turbulence we considered $\alpha= 2.064$, $\beta = 1.342$.

In Fig. 2, simulation and theoretical results are displayed for moderate and strong turbulence conditions when equal power allocation is performed across different HARQ rounds. It becomes obvious from the figure that analytical results coincide with simulations for the HARQ-CC scheme; thus validating theoretical analysis. On the other hand, regarding HARQ-IR scheme, there is a slight error between simulation and analytical results which was emerged from the approximation of OP. It's visible that the approximate expression forms a tight upper-bound across the whole range of considered SNR values. In addtion, it can be observed that HARQ-IR performs better from HARQ-CC in terms of OP for different fading conditions. Finally, as a benchmark, analytical results for the cases of $J=1$ (without HARQ) and $J=10$ are also included in the same figure, revealing the effectiveness of HARQ schemes as the number of retransmissions increases as well as the fast convergence to the extreme case.


\begin{figure}[h]
\centering
\includegraphics[width=0.65\linewidth]{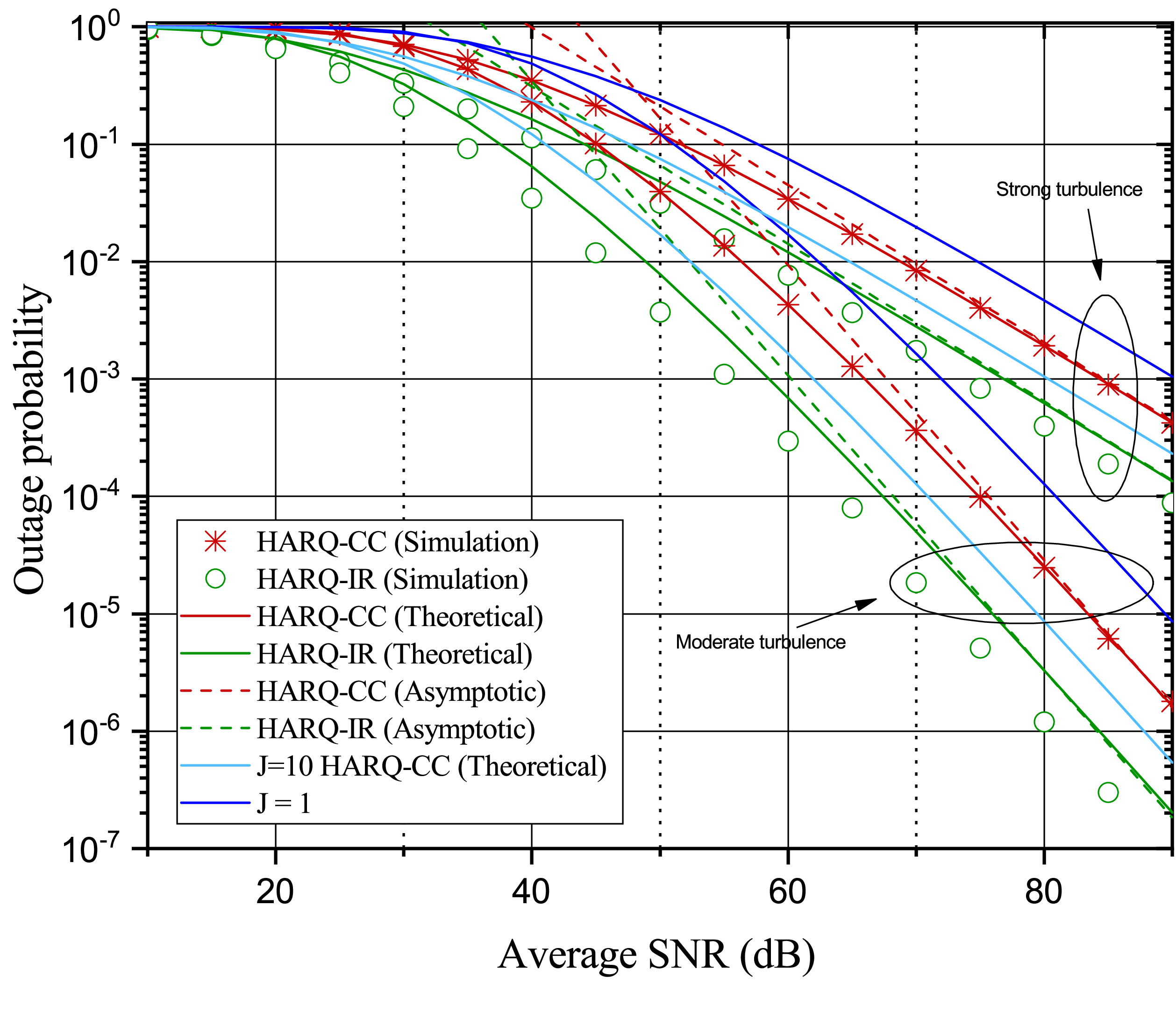}
\vspace{-0.4cm}
\caption{OP for different turbulent conditions, $R \!=\! 2$ bits/s/Hz, $P = P_{\text{max}}$ }  \label{fig:fig1}
\end{figure}

\begin{figure}[h]
\vspace{-0.3cm}
\centering
\includegraphics[width=0.65\linewidth]{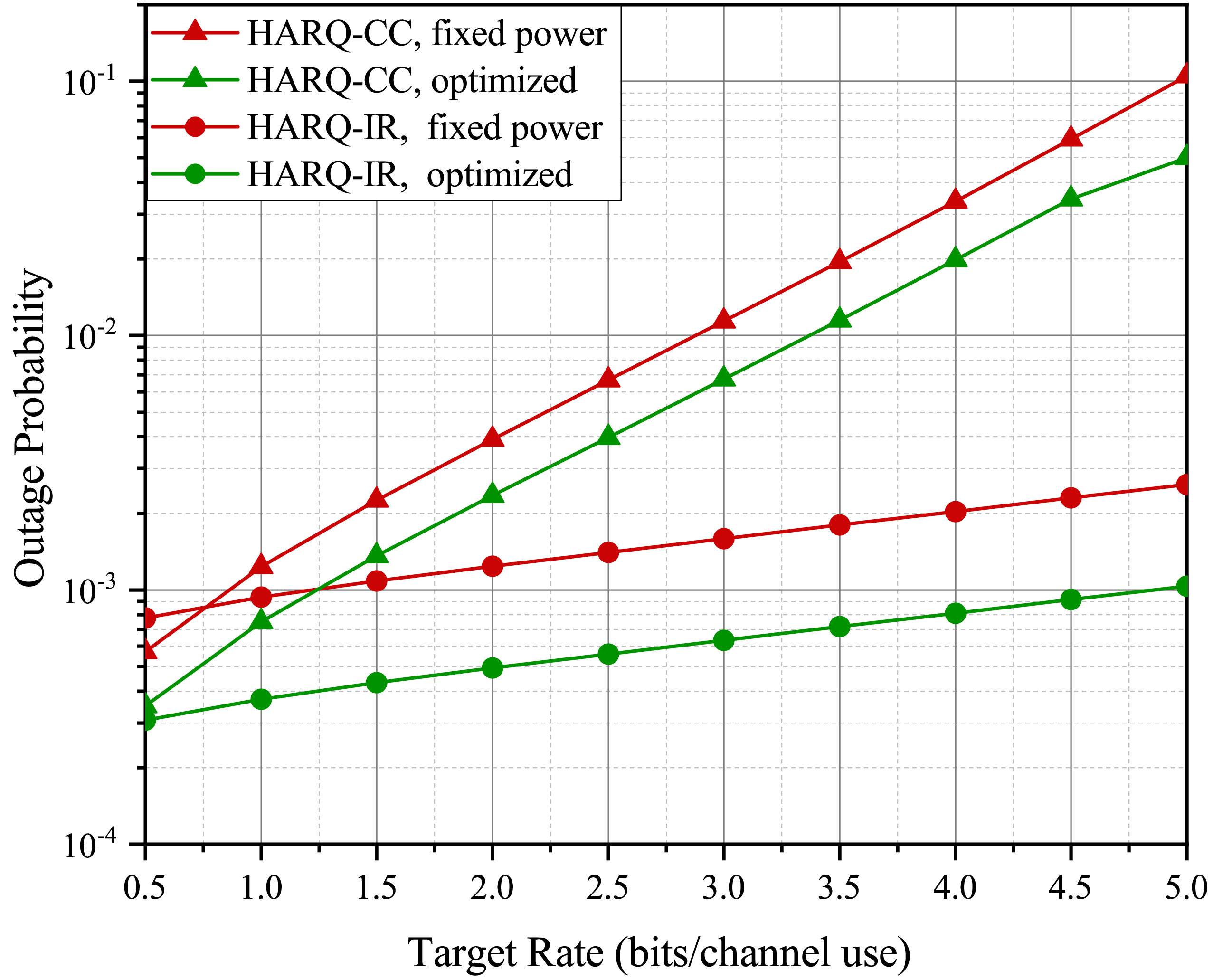}
\caption{OP vs the target rate, $\bar{\gamma}=60$dB.}  \label{fig:fig1}
\end{figure}

\begin{figure}[!h]
\centering
\includegraphics[width=0.65\linewidth]{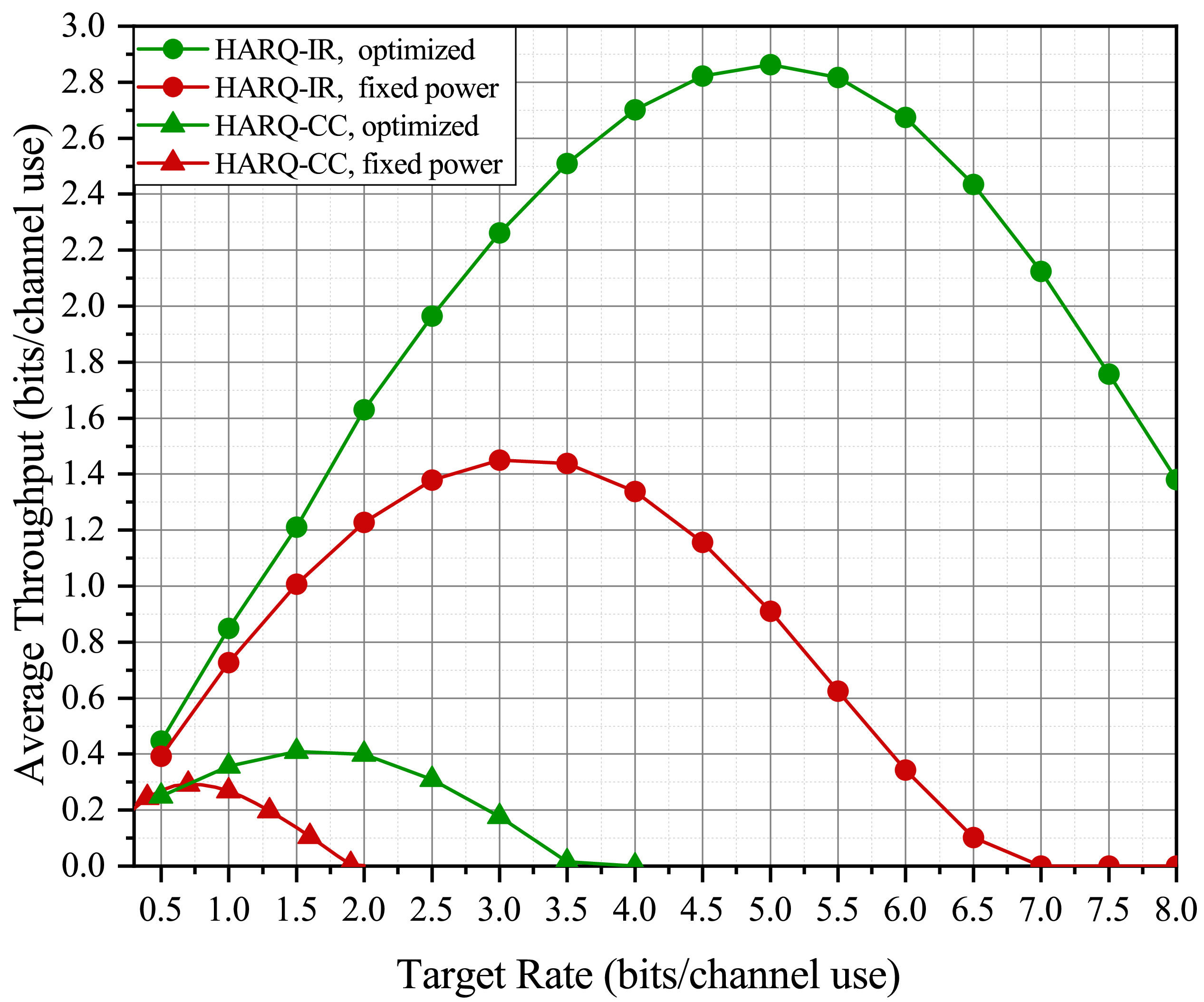}
\caption{Average Throughput vs the target rate, $\bar{\gamma}=30$dB.}  \label{fig:fig1}
\vspace{-0.4cm}
\end{figure}

\begin{figure*}[!h]
\small
	\begin{align}
		\mathbb{P}_{\text {out},j}^{\mathrm{\textsc{cc}}} &= \frac{\Gamma\left(\alpha-\xi^2\right)\Gamma\left(\beta - \xi^2 \right)}{\Gamma\left(\alpha\right)\Gamma\left(\beta\right)} \left(\frac{V_R}{\sum_{i=1}^{j} P_i^2}\right)^{\xi^2/2} \!+\! \frac{\Gamma\left(\beta-\alpha\right)}{\left(1-\alpha/\xi^2\right)\Gamma\left(\alpha+1\right)\Gamma\left(\beta\right)}  \left(\frac{V_R}{\sum_{i=1}^{j} P_i^2}\right)^{\alpha/2}
		\nonumber\\ &\times { }_2 F_3\left(\alpha,\alpha-\xi^2; 1+\alpha-\xi^2,1+\alpha-\beta,1+\alpha ; \sqrt{\frac{V_R}{\sum_{i=1}^{j} P_i^2}} \right) + \frac{\Gamma\left(\alpha-\beta\right)}{\left(1-\beta/\xi^2\right)\Gamma\left(\beta+1\right)\Gamma(\alpha)}
		\left(\frac{V_R}{\sum_{i=1}^{j} P_i^2}\right)^{\beta/2}	
		\nonumber\\ &\times { }_2 F_3\left(\beta,\beta-\xi^2; 1+\beta-\xi^2,1+\beta-\alpha,1+\beta ; \sqrt{\frac{V_R}{\sum_{i=1}^{j} P_i^2}} \right), \, \text{with} \;\:  V_R = \alpha^2 \beta^2 \left(2^{2R}-1\right)/c\bar{\gamma}
		\label{eq:28}
	\end{align}
	\hrulefill
    \vspace{-0.4cm}
\end{figure*}

In Fig. 3, the OP is plotted against different values of the target rate $R$ for moderate turbulent conditions. As benchmarks, the HARQ-CC and the HARQ-IR schemes with constant power during each round $j, \forall j \in \{1,...,J\}$, namely $P_j=P_0/J$, are also depicted in the figure. It can be seen that by proper power allocation, as presented in section \ref{sec:HARQ}, significant improvement in terms of OP is achieved. Specifically, the optimized HARQ-CC outage performance is three times better compared to its non-optimized version, while the optimized HARQ-IR is about five times better than its non-optimized counterpart. For low $R$, retransmissions are barely exploited; therefore, the performance of both protocols is similar. However, HARQ-IR completely dominates the HARQ-CC protocol for greater values of rates $R$, due to increased complexity.



In Fig. 4, the optimized average throughput, as given in section \ref{sec:HARQ}, is illustrated 
for moderate turbulence. For the HARQ-CC protocol, it is observed that its optimized version offers a peak average throughput of 0.4 bits/channel while its non-optimized counterpart achieves 0.3 bits/channel peak average throughput. Furthermore, the optimized HARQ-IR is shown to be doubled compared to the fixed power HARQ-IR benchmark. In particular, the optimized power HARQ-IR scheme achieves 2.8 bits/channel use, while its fixed power counterpart achieves 1.4 bits/channel use. Thus, it is evident that the power allocation in HARQ schemes greatly improve the efficiency of the FSO system.



\section{Conclusions}\label{sec:Conclusions}
In this paper we explored the power efficient and reliable employment of the HARQ protocol for FSO communications.
In particular, two  HARQ schemes were examined, HARQ-CC and HARQ-IR. The outage performance of both HARQ schemes was investigated and closed-form analytical expressions were provided. Furthermore, the problem of minimizing the OP of the considered schemes under average and maximum optical power constraints was addressed. Power allocation schemes were developed, which distribute the optical power across the retransmission rounds and offer reduced power consumption as well as improve the throughput of the FSO systems. Simulation results validated the presented analysis and illustrated the improvement of the proposed HARQ schemes, in terms of both OP and throughput, compared to the fixed transmit power HARQ schemes. As a future study, FSO's performance can improved by optimally adjsting the code rate of the HARQ-IR protocol across different retransmission rounds.
\section*{Acknowledgement}
{This work has received funding from the European Unions Horizon-JU- SNS-2022 research and innovation programme under grant agreement No. 101096456 (NANCY). }
\section*{Appendix}
In the case of block-fading channel across all HARQ-CC rounds, the OP can be expanded using the \cite[(9.303)]{B:grads} as \eqref{eq:28}, where ${ }_p F_q\left(\cdot ; \cdot ; \cdot\right)$ is the Generalized hypergeometric function. 
In high-SNR region we can see that $V_R \rightarrow 0$. Then, using the known property of Generalized hypergeometric function $\lim_{z\rightarrow 0^+}{ }_p F_q\left(a_1, \ldots, a_p ; b_1, \ldots, b_q ; z\right) = 1$, we can write the asymptotic OP as the sum of three terms. The most dominant term is decided based on the values of $\xi^2,\alpha,\beta$. So, if $\xi^2 < \min\left\{\alpha,\beta\right\}$, then the OP can be approximated as

\begin{equation}
\small
	\mathbb{P}_{\text {out},j}^{\mathrm{\textsc{cc}}} \approx \frac{\Gamma\left(\alpha-\xi^2\right)\Gamma\left(\beta - \xi^2 \right)}{\Gamma\left(\alpha\right)\Gamma\left(\beta\right)} \left(\frac{V_R}{\sum_{i=1}^{j} P_i^2}\right)^{\xi^2/2}.
\end{equation}

if $\xi^2 > \min\left\{\alpha,\beta\right\}$  the dominant term comes from the minimum of $\alpha$ and $\beta$ and is given by

\begin{equation}
\small
\begin{aligned}
	\mathbb{P}_{\text {out},j}^{\mathrm{\textsc{cc}}}& \!\approx\! \frac{\Gamma\left(\lvert\beta-\alpha\rvert\right)}{\left(1-\min\{\alpha,\beta\}/\xi^2\right)\Gamma\left(\min\{\alpha,\beta\}+1\right)\Gamma\left(\max\{\alpha,\beta\}\right)}
	\nonumber\\ & \times \left(\frac{V_R}{\sum_{i=1}^{j} P_i^2}\right)^{\min\{\alpha,\beta\}/2}.
 \end{aligned}
\end{equation}









\vspace{-0.3cm}
\linespread{0.9}\selectfont
\bibliographystyle{IEEEtran}
\bibliography{IEEEabrv,References}
\end{document}